\documentclass[conference]{IEEEtran}
\IEEEoverridecommandlockouts
\usepackage{cite}
\usepackage{amsmath,amssymb,amsfonts}
\usepackage{algorithmic}
\usepackage{graphicx}
\usepackage{textcomp}
\usepackage{xcolor}
\def\BibTeX{{\rm B\kern-.05em{\sc i\kern-.025em b}\kern-.08em
    T\kern-.1667em\lower.7ex\hbox{E}\kern-.125emX}}
\begin{document}

\title{Toward Fully Neuromorphic Receivers for Ultra-Power Efficient Communications
}

\author{
\IEEEauthorblockN{George N. Katsaros and Konstantinos Nikitopoulos}
\IEEEauthorblockA{
Wireless Systems Lab, 5G \& 6G Innovation Center\\
Institute for Communication Systems, University of Surrey, United Kingdom
}
}

% \author{\IEEEauthorblockN{George N. Katsaros}
% \IEEEauthorblockA{\textit{(COMMON + WSL)5G \& 6G Innovation Center} \\
% \textit{University of Surrey}\\
% United Kingdom}
% \and
% \IEEEauthorblockN{Konstantinos Nikitopoulos}
% \IEEEauthorblockA{\textit{5G \& 6G Innovation Center} \\
% \textit{University of Surrey}\\
% United Kingdom}
% }

\maketitle

\begin{abstract}
Neuromorphic computing, inspired by biological neural systems, has emerged as a promising approach for ultra-energy-efficient data processing by leveraging analog neuron structures and spike-based computation. However, its application in communication systems remains largely unexplored, with existing efforts mainly focused on mapping isolated communication algorithms onto spiking networks, often accompanied by substantial, traditional computational overhead due to transformations required to adapt problems to the spiking paradigm.
In this work, we take a fundamentally different route and, for the first time, propose a fully neuromorphic communication receiver by applying neuromorphic principles directly in the analog domain from the very start of the receiver processing chain. Specifically, we examine a simple transmission scenario: a BPSK receiver with repetition coding, and show that we can achieve joint detection and decoding entirely through spiking signals. Our approach demonstrates error-rate performance gains over conventional digital realizations with power consumption on the order of microwatts, comparable with a single very low-resolution Analog-to-Digital Converter (ADC) utilized in digital receivers. To maintain performance under varying noise conditions, we also introduce a novel noise-tracking mechanism that dynamically adjusts neural parameters during transmission. Finally, we discuss the key challenges and directions toward ultra-efficient neuromorphic transceivers.

\end{abstract}


\section{Introduction}
Neuromorphic computing has emerged as a compelling approach for ultra energy-efficient data processing, with potential power reductions of several orders of magnitude compared to conventional systems \cite{schuman_opportunities_2022}. 
Drawing inspiration from the event-driven computation and the high energy efficiency observed in biological neurons, neuromorphic circuits employ fundamentally analog structures and spike-based signaling, representing a significant departure from conventional Von Neumann architectures. Apart from ultra-low power consumption, neuromorphic computing promises substantial benefits, including the elimination of memory bottlenecks and high scalability achieved through extensive replication of simple neuron-based building blocks, thus enabling massive parallelism \cite{schuman_survey_2017}.

Despite these promising attributes, the potential of neuromorphic computing in communication systems remains largely unexplored. Current applications predominantly focus on replacing existing artificial neural network implementations~\cite{song_neuromorphic_2024} with spiking networks, leaving a broader range of computationally intensive problems beyond the scope of machine learning open. Efforts to apply neuromorphic principles to non-machine-learning tasks for physical-layer (PHY) processing, such as different Multi-User Multiple-Input Multiple-Output (MU-MIMO) detection algorithms, have been made~\cite{katsaros_neuromimo_2024,katsaros_towards_2024} but typically examined in isolation of a complete system. Although these approaches demonstrate potential gains over the traditional ones, they commonly rely on mathematically reformulating the original problem into appropriate forms that a specific spiking network topology can solve (e.g., into a Quadratic Unconstrained Binary Optimization (QUBO) representation). This preprocessing step requires execution on a traditional computing platform and can significantly constrain the full range of advantages that a complete neuromorphic processing pipeline could otherwise achieve.

In this work, we take a different approach. Rather than focusing on mapping complex communication algorithms in isolation, we explore for the first time how to construct a fully neuromorphic receiver. We examine the example case of a simple Binary Phase Shift Keying (BPSK) receiver with repetition coding under a narrow transmission bandwidth and apply neuromorphic principles from the beginning of the processing chain directly in the analog domain. Using this approach, we achieve joint detection and decoding via spiking signals that, as we later show, can offer tangible improvements in both error-rate performance and power consumption compared to their equivalent digital-based counterpart.
In addition, we discuss the challenges facing future neuromorphic receivers, including the processing speed limitations (spike rates) of current analog-based neuron designs and the importance of selecting robust neural parameters capable of handling varying noise conditions. In this direction, also for the first time, we introduce a noise-tracking mechanism that dynamically adjusts specific neural parameters (i.e., the corresponding neuron spiking thresholds) in response to the estimated noise to maintain the error-rate performance.
Finally, our initial power consumption estimates indicate that the proposed neuromorphic receiver can operate at the order of microwatts ($\mu W$), typically lower than the power consumed by a single low-resolution ADC required for the equivalent digital implementation.

The remainder of the paper is structured as follows. Sec.~\ref{sec:bg} introduces the fundamentals of neuromorphic processing, presents our system model, and describes the key ideas behind our mapping approach. Sec.~\ref{sec:system_design} presents the proposed neuromorphic architectures, details the parameter selection for the neurons, and introduces the noise-tracking mechanism. Finally, Sec.~\ref{sec:results} presents our experimental results and comparisons with the equivalent digital implementation and discusses the remaining open challenges.

\section{Neuromorphic Symbol Detection}\label{sec:bg}
\subsection{Fundamentals of Neuromorphic Processing}\label{ssec_bg1}

The fundamental computing units in neuromorphic systems are the neurons. Neurons accumulate charge from either external stimuli or incoming spikes from other neurons. When this charge reaches a threshold, the neuron fires, transmitting a spike to connected neurons via synapses.
Synapses link neurons and are characterized by a synaptic weight, which may be excitatory or inhibitory, and optionally a delay that introduces signal propagation time. Neurons may also exhibit leakage, where charge dissipates over time, and a refractory period after firing, during which they cannot spike again.
Neuromorphic systems can be implemented using analog circuits that emulate continuous-time dynamics or digital ones that approximate them discretely. These systems typically offer a configurable interconnection fabric and exhibit high energy efficiency due to their event-driven, massively parallel architecture \cite{schuman_opportunities_2022,truenorth}. Such features make them especially suitable for PHY-layer algorithms with inherent parallelism, such as those spanning subcarriers or symbols.

This work employs a standard leaky integrate-and-fire (LIF) neuron model, widely adopted in neuromorphic hardware \cite{young_review_2019,loihi,truenorth}. Its behavior is captured by the differential equation: 
\begin{equation} \tau_m \frac{dV(t)}{dt} = -(V(t)-V_{rest}) +I(t) \label{eq_lif} \end{equation}
where $V(t)$ is the membrane potential, $V_{rest}$ is the resting potential, $\tau_m$ is the membrane time constant governing leakage, and $I(t)$ is the synaptic or external input. When $V(t)$ reaches a threshold $V_{th}$, the neuron fires and resets to $V_{rest}$.
In the following sections, we introduce the proposed LIF-based architecture for energy-efficient joint detection-decoding.

\subsection{Analog Symbol Detection in Repetition-Coded BPSK}

Consider a repetition-coded BPSK system, where each transmitted bit $s_m$ takes values in $\{-1,+1\}$ and is repeated $n$ times. Let the received signals be $r_i$ for $i = 1, 2, \dots, n$. Assuming independent Gaussian noise with variance $\sigma^2$ and equally likely transmitted symbols, the maximum-likelihood (ML) detector selects the symbol $s_m$ that maximizes the conditional likelihood:
\begin{equation}
  \hat{s}_m
  \;=\; \arg \max_{s_m \in \{-1,+1\}}
    P\bigl(r_1,\,r_2,\dots,r_n \,\bigm|\,
    s_m\bigr)
\end{equation}
For an additive white Gaussian noise (AWGN) channel, each $r_i$ is drawn from a normal distribution $\mathcal{N}(s_m,\sigma^2)$. Hence,
\begin{equation}
  P\bigl(r_1,\dots,r_n \,\bigm|\,
    s_m\bigr)
  \;=\;
  \prod_{i=1}^n
  \frac{1}{\sqrt{2\pi}\,\sigma}
  \exp\!\Bigl(
    -\tfrac{|\,r_i - s_m\,|^2}{2\sigma^2}
  \Bigr)
\end{equation}
Taking the log-likelihood ratio $\lambda$ of the hypotheses $s_m=+1$ versus $s_m=-1$ gives:
\begin{align}
  \lambda
  &= \ln\!
    \biggl(
      \frac{P(r_1,\dots,r_n \mid s_m=+1)}
           {P(r_1,\dots,r_n \mid s_m=-1)}
    \biggr)
  \nonumber\\[6pt]
  &= \sum_{i=1}^n
     \ln\!
     \Bigl(
       \exp\!\Bigl(
         -\tfrac{|\,r_i - (+1)\,|^2}{2\sigma^2}
       \Bigr)
       \,\Big/\,
       \exp\!\Bigl(
         -\tfrac{|\,r_i - (-1)\,|^2}{2\sigma^2}
       \Bigr)
     \Bigr)
  \nonumber\\
  &= \sum_{i=1}^n
     \Bigl[
       -\tfrac{(r_i - 1)^2}{2\sigma^2}
       \;+\;
       \tfrac{(r_i + 1)^2}{2\sigma^2}
     \Bigr]
  \;=\;
  \frac{2}{\sigma^2}\,\sum_{i=1}^n r_i\,
  \label{eq:LLR}
\end{align}
The decision rule then reduces to comparing $\lambda$ with zero. Since $\lambda$ is proportional to the sum of the received samples $\sum_{i=1}^n r_i$, we obtain:
\begin{equation}
  \hat{s}_m 
  \;=\;
  \mathrm{sign}\!\bigl(\lambda\bigr)
  \;=\;
  \mathrm{sign}\!\Bigl(\sum_{i=1}^n r_i\Bigr)
  \label{eq:decision}
\end{equation}
Thus, the ML decision for repetition-coded BPSK can be performed in the analog domain by simply summing the $n$ received signals and taking the sign of the result.

\subsection{Mapping BPSK Detection to Spiking Neurons}

In a neuromorphic computing context, the received analog signal $r_i$ doesn't need to be sampled or digitized. Instead, it can be directly translated into a spiking signal whose firing rate (or spike count within a fixed time window) reflects the amplitude of $r_i$. In this way, the ``soft'' information, typically extracted via amplitude-sensitive ADCs, is preserved in the number of spikes produced by a LIF neuron or an equivalent spiking circuit.

Suppose each repeated transmission of a symbol contributes to a continuous-time waveform that is fed into a single encoding LIF neuron. This neuron emits spikes at a rate $\rho(r(t))$, where $\rho(\cdot)$ is a saturating, approximately linear function of the instantaneous input amplitude. Over the full symbol duration $T$ (including all repetitions), the total number of emitted spikes is denoted by $N = \rho(r)\,T$, and implicitly encodes the joint information from all repetitions.
To perform joint detection, analogous to the summation of received samples in Eq.~\eqref{eq:decision}, the resulting spike train is accumulated by a second neuron, referred to as the detection neuron. This neuron integrates spikes over the interval $T$ and fires whenever the total spike count exceeds a predefined threshold, as described in Section~\ref{ssec_bg1}. The decision rule can then be translated to:
\begin{equation}
  \hat{s}_m \;=\;
  \begin{cases}
    +1, & \text{if the detection-neuron fires within } T, \\[6pt]
    -1, & \text{otherwise}.
  \end{cases}
\end{equation}
In this way, the neuromorphic architecture directly mirrors the continuous-time summation of $r_i$ without requiring analog-to-digital conversion. 
Instead, the signal amplitude is naturally embedded in the spiking dynamics and combined in the analog domain via spike-based accumulation. 
Building on this conceptual foundation, the following Section describes the system architecture that enables the practical realization of the neuromorphic detector-decoder in hardware.

\begin{figure*}[ht]
    \centering
    \includegraphics[height=6cm]{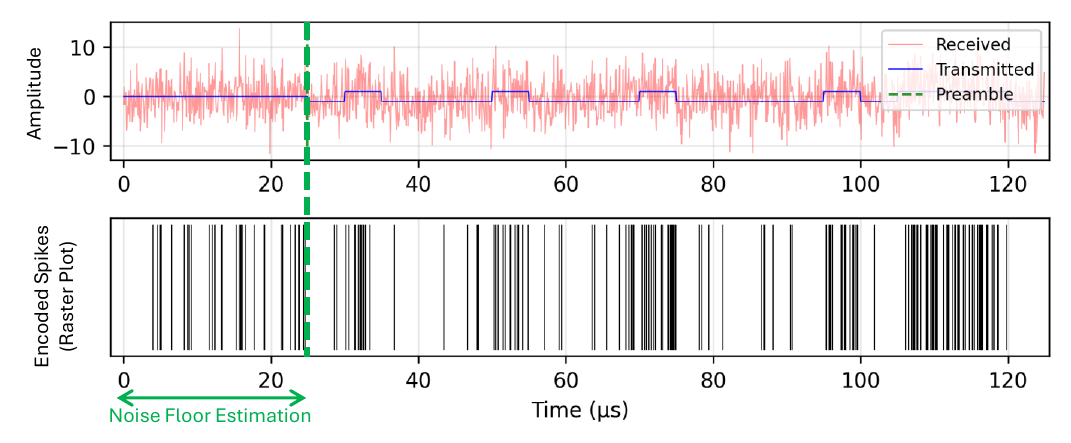}
    \caption{Time-domain waveforms and corresponding spike raster for a BPSK transmission with a noise-estimation preamble (green dashed region). The top panel shows the transmitted (blue) and received (red) signals, while the bottom panel illustrates the spike-encoded representation.}
    \label{fig:raster}
\end{figure*}

\section{System Architecture}\label{sec:system_design}

\begin{figure}[!t]
    \centering
    \includegraphics[width=\linewidth]{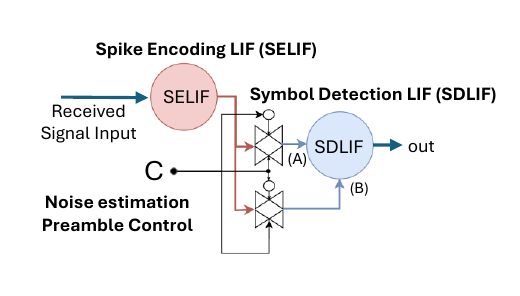}
    \caption{Simplified architecture of the proposed neuromorphic detection system. The Spike Encoding LIF (SELIF) neuron provides spike inputs via a transmission gate, (A) to the Symbol Detection LIF (SDLIF) for data detection, while (B) noise-induced spikes from the preamble interval to dynamically adjust the SDLIF threshold.}
    \label{fig:system_arch}
\end{figure}
The proposed neuromorphic detector performs symbol detection directly in the analog domain using a cascade of just two LIF neurons. The system targets BPSK-modulated communication with configurable repetition coding of rate $1/n$. The proposed architecture is illustrated in Fig.~\ref{fig:system_arch}.
First, a \textit{Spike Encoding LIF neuron} (SELIF) converts the analog baseband signal into a discrete-time spike train, where the number of spikes within a symbol interval is proportional to the received signal amplitude. This is achieved by using the baseband analog signal directly to drive increases and decreases of the SELIF membrane potential.
As an example, Fig.~\ref{fig:raster} (bottom) shows the produced spike train for a BPSK transmission of 20 symbols with $T_{\text{sym}} = 5\,\mu\text{s}$ at an SNR of $4$\,dB and without repetition coding. When the input (Fig.~\ref{fig:raster} (top)) corresponds to a “+1” symbol, the neuron emits spikes at a high rate. In contrast, for a “--1” symbol, the spike rate decreases substantially.
As mentioned in Section~\ref{sec:bg}, rather than summing analog samples $r_i$, we accumulate the spikes emitted in response to those signals. With a repetition code applied, the key insight is that spike counts aggregated over $n$ symbol repetitions ($n \cdot T_{\text{sym}}$) preserve soft information without requiring high-resolution sampling or digital reconstruction.

The symbol detection logic is implemented via a second cascaded neuron, the \textit{Symbol Detection LIF} (SDLIF) neuron, which integrates the encoded spikes. Each incoming spike increases the SDLIF membrane potential by a small increment $\mathrm{d}V_m$. Hence, the detection rule becomes:

\begin{equation}
    \hat{s}_m = 
    \begin{cases}
        +1, & \text{if } N_{\text{spikes}} \cdot \mathrm{d}V_m \;\geq\; \theta, \\[4pt]
        -1, & \text{otherwise,}
    \end{cases}
\end{equation}
where $N_{\text{spikes}}$ is the total spike count over the $n$ repetitions of a symbol.
A natural choice for $\theta$ is a fixed threshold based on the maximum expected spike count at high SNR. For example, if the SELIF has a membrane time constant $\tau_m = 0.5\,\mu\mathrm{s}$ and threshold voltage $v_{\mathrm{th}} = 0.5$, then the theoretical interspike interval follows from Eq.~\eqref{eq_lif} at $V(t_{\text{spike}}) = v_{\mathrm{th}}$:
\begin{equation}
    t_{\text{spike}} 
    = \tau_m \ln\biggl(\frac{1}{1 - v_{\mathrm{th}}}\biggr) 
    \approx 0.3465\,\mu\mathrm{s}
\end{equation}
With a symbol duration of $T_{\text{sym}} = 5\,\mu\mathrm{s}$, this yields an ideal spike count of approximately 14.4 spikes for a “+1” input. Consequently, a convenient static threshold for the SDLIF can be set to about half of this nominal count to distinguish between “–1” and “+1” decisions, i.e., 
$\theta \approx 7 n \cdot dV_m$, 
where $n$ is the repetition factor.

However, in practice and as shown later in Section~\ref{sec:results}, the static threshold approach can fail under low-SNR conditions. In such scenarios, noise alone may trigger spiking during the transmission of “--1” symbols, leading to false positives. To address this, we introduce a noise-adaptive thresholding mechanism. Specifically, we insert a fixed-duration preamble of silence before each transmission frame. During this silent interval, the SELIF neuron is driven only by noise, and the spikes emitted provide an estimate of the noise-induced firing rate.
Let $N_{\text{noise}}$ denote the total number of spikes generated during a preamble interval spanning $n_{\text{symbols}}$ symbol durations. Each preamble-induced spike incrementally raises the detection threshold voltage of the SDLIF neuron by a fixed increment $dV_{\mathrm{n}}$, determined by:
\begin{equation}
    dV_{\mathrm{n}} = \frac{ n \cdot dV_m}{n_{\text{symbols}}}
\end{equation}
where $n$ is the repetition factor and $dV_m$ is the nominal membrane voltage increment per spike. Consequently, the adaptive threshold voltage for the SDLIF neuron is given by:
\begin{equation}
    \theta_{\text{adaptive}} 
    = N_{\text{noise}} \cdot dV_{\mathrm{n}}
\end{equation}
The proposed approach eliminates the need for explicit digital computation, instead leveraging the intrinsic physical properties of the neuron circuit to dynamically track the noise floor. The spikes generated during the preamble interval adjust the SDLIF neuron's firing threshold (via SDLIF Neuron Interface-B Fig.~\ref{fig:system_arch}), whereas the data-induced spikes increase the neuron's membrane potential (via Interface-A Fig.~\ref{fig:system_arch}). Finally, because there is no active leakage path during threshold integration, the accumulated threshold voltage remains stable until a new noise estimation phase is initiated.

\section{Results and Discussion}\label{sec:results}

\begin{figure}
    \centering
    \includegraphics[width=\linewidth]{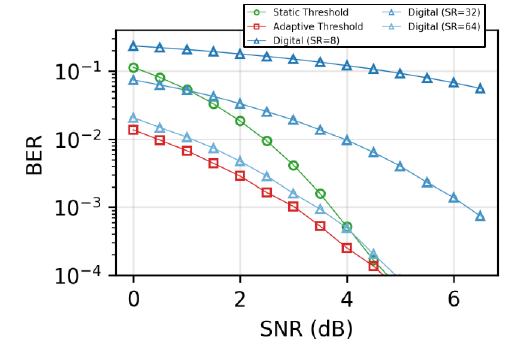}
    \caption{Bit error rate (BER) comparison between the proposed neuromorphic detectors and conventional digital operating at various 2-bit ADC sampling rates (SR) in samples per symbol, with 1/3 repetition coding.} 
    \label{fig:adc_comparison}
\end{figure}

% Experiment two different silence period duration (noise Est accuracy)
\begin{figure}
    \centering
    \includegraphics[width=\linewidth]{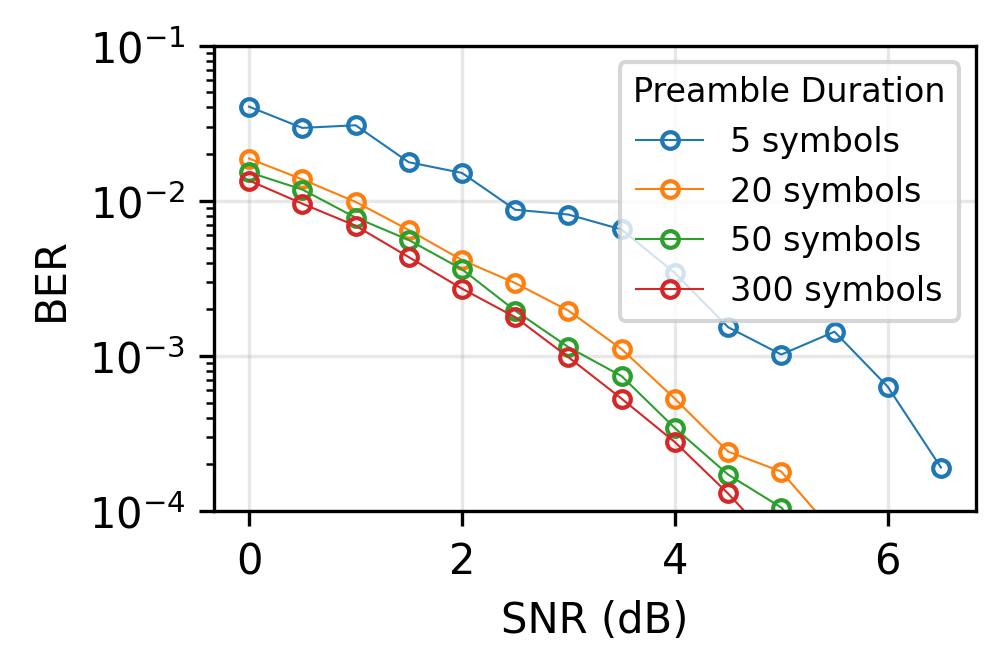}
    \caption{Bit error rate (BER) performance of the proposed adaptive-threshold neuromorphic detector for various preamble durations with 1/3 repetition coding.} 
    \label{fig:preamble_sweep}
\end{figure}

We simulate the proposed detection-decoding system utilizing a symbol duration $T_{\text{sym}} = 5\,\mu\text{s}$ and repetition code of rate $1/3$. We perform a total of 1 million BPSK transmissions per signal-to-noise ratio (SNR) point over a range of SNR values. All signal processing simulations are implemented in Python. 
For our first experiment, three detector-decoder configurations are tested as discussed also in Sec. \ref{sec:system_design}: (\emph{i}) the neuromorphic detector-decoder employing a static threshold based on high-SNR spike counts, (\emph{ii}) the neuromorphic detector-decoder with a 300-symbol silent preamble to adaptively estimate noise-induced spikes, and (\emph{iii}) the digital processing equivalent with a 2-bit analog-to-digital converter (ADC) sampling at $64$ points per symbol.

As shown in Fig.~\ref{fig:adc_comparison}, the spiking neuromorphic detector can efficiently achieve practical BER performance, outperforming the ADC-based digital detector when the adaptive threshold mechanism is employed.
In addition, to enable a more equitable comparison with the digital detector, we examine the equivalent digital detection-decoding schemes employing ADCs with varying sampling rates. The sampling rate, expressed in samples per symbol (i.e, 8, 32, 64), directly influences the power consumption of the ADC circuitry. As shown, the performance advantage of the spiking detectors becomes increasingly pronounced as lower-rate (and thus lower-power) ADCs are considered, highlighting the efficiency of the proposed analog spiking approach under power-constrained scenarios.
Importantly, as shown in Fig.~\ref{fig:adc_comparison}, the adaptive threshold approach offers substantial gains compared to the static threshold neuromorphic detector, particularly in the low-SNR regime, effectively mitigating false detections. 

However, the noise-adaptability comes at the cost of preamble overhead. Fig.~\ref{fig:preamble_sweep} quantifies the impact of varying preamble duration on detection performance for preamble durations of 5, 20, 50, and 300 symbols. As shown, performance improves rapidly with increasing preamble length, but beyond 50 symbols, the gains become marginal. This indicates that effective noise adaptation can be achieved with relatively short preambles, limiting the overhead while preserving detection accuracy.

\subsection{The Power Gains of the Neuromorphic Approach}

To better estimate the power consumption of the proposed neuromorphic front-end, we consider its core components: two leaky integrate-and-fire (LIF) neurons and a transmission gate. In order to support symbol durations on the order of $5\,\mu\mathrm{s}$ while producing multiple spikes per symbol, the LIF neuron is configured with a relatively short membrane time constant, such as $\tau_m = 0.5\,\mu\mathrm{s}$. At this timescale, the neuron repeatedly charges and discharges its membrane capacitor $C_m$ up to a threshold voltage $v_{\text{th}}$, with each spike event consuming energy approximately equal to $\frac{1}{2}\,C_m\,v_{\text{th}}^2$, typically falling in the range of $1$--$10\,\mathrm{pJ}$~\cite{subbulakshmi_radhakrishnan_biomimetic_2021,mushtaq_energy_2024}. 
Assuming an average spike count of 14 spikes per symbol when transmitting a ``1'' at high SNR as discussed in Sec. \ref{sec:system_design}, the corresponding energy per symbol is approximately $14 \times 5\,\mathrm{pJ} = 70\,\mathrm{pJ}$. Averaged over the symbol duration ($T_{\mathrm{sym}} = 5\,\mu\mathrm{s}$), this corresponds to an estimated dynamic power consumption of $14\,\mu\mathrm{W}$ per LIF neuron operating at a symbol rate of $200\,\mathrm{kSymbols/s}$. The accompanying transmission gate, assuming switching at approximately $2\,\mathrm{kHz}$ to alternate between preamble and data transmission, contributes only a few picowatts.

For comparison, typical low-resolution ADCs (2 bits) operating at sampling rates of a few MS/s (i.e., 12.8~MS/s for 64 samples per symbol as of Fig.~\ref{fig:adc_comparison}) consume on the order of $0.5-1\,\mathrm{mW}$ with higher-resolution, higher-speed ADCs exhibiting significantly greater power consumption~\cite{nikitopoulos_digilogue_2021}. These preliminary estimates suggest that the proposed spiking front-end can achieve symbol-level detection-decoding at lower power levels, compared to equivalent digital approaches, while also improving on the error-rate performance. It is important to note here that the comparison only involves the ADC and not the subsequent required logic for the signal processing; therefore, the actual gains of the neuromorphic approach can be significantly higher. More importantly, studies~\cite{sourikopoulos_4-fjspike_2017,zhu_optical_2022} have reported LIF implementations operating at even lower energy levels, down to the femtojoule range per spike.

\section{Conclusions and Future Work}\label{sec:conc}
In this work, we presented a first fully analog neuromorphic joint detection-decoding architecture. By leveraging LIF neurons to encode and accumulate amplitude information directly in the analog domain, without requiring ADCs or digital computation, the proposed approach offers a highly efficient solution for repetition-coded systems. Initial results demonstrate improved error-rate performance compared to digital processing equivalents, particularly when combined with the proposed noise-tracking mechanism. Furthermore, the system exhibits power consumption at the order of microwatts, typically less than just a single ADC required by the equivalent digital processing detection-decoder.

Future extensions will focus on generalizing the mathematical framework and the hardware architecture to support denser constellations, such as multi-level Pulse Amplitude Modulation (PAM) or Quadrature Amplitude Modulation (QAM). This will require the deployment of additional detection neurons and the development of multi-threshold mechanisms to distinguish between multiple symbol levels reliably. 
Moreover, supporting larger transmission bandwidths in the analog domain requires LIF neurons capable of operating with very short inter-spike intervals. While this remains a limitation in most bio-inspired implementations, technologies such as photonic circuits have demonstrated neuromorphic processing at gigahertz-scale spiking rates without compromising energy efficiency \cite{owen-newns_photonic_2023}. Nonetheless, the inherent scalability of neuromorphic architectures enables massively parallel implementations, allowing real-time performance to be achieved even with lower individual spiking rates.
Finnaly, future work will aim to extend the architecture toward multi-user MIMO scenarios directly in the analog domain for next-generation ultra-power efficient wireless tranceivers.

\bibliographystyle{ieeetr} 
\bibliography{refs}

\end{document}